\documentclass[aps,prb,twocolumn,superscriptaddress]{revtex4-2}

\usepackage{graphicx} 
\usepackage{graphics}
\usepackage{epsfig}
\usepackage{appendix}
\usepackage{amsmath, amssymb, braket}
\usepackage{dirtytalk}
\usepackage{hyperref, listings}
\usepackage{svg}
\usepackage{comment}
\usepackage{float}

\begin{document}

\title{Inducing cooperative dynamics in an inhomogeneous many-emitter ensemble with a pulse  sequence}

\author{V. P. Iglesias-Cardinale}
\email{vpiglesi@buffalo.edu}
\affiliation{Department of Physics, University at Buffalo SUNY, Buffalo, New York 14260, USA}

\author{Shreekanth S. Yuvarajan} 
\affiliation{Department of Physics, University at Buffalo SUNY, Buffalo, New York 14260, USA}

\author{Herbert F. Fotso}
\email{hffotso@buffalo.edu}
\affiliation{Department of Physics, University at Buffalo SUNY, Buffalo, New York 14260, USA}

\begin{abstract}
\noindent Driving a many-particle quantum system with an appropriately tuned external field can give rise to features in its nonequilibrium dynamics that are absent in the equilibrium counterpart. Cooperative dynamics in a dense ensemble of quantum emitters manifested through superradiance, a burst in the emission intensity as a function of time, is typically suppressed by spectral disparities among emitters in the ensemble. We demonstrate the enhancement of cooperative dynamics in a spectrally inhomogeneous ensemble of quantum emitters when it is driven by a periodic pulse sequence. Using both a semi-analytical mean-field solution and a digital quantum simulation, we examine the coherence of the pulse-driven ensemble and the intensity of the emitted radiation as a function of the spectral disparity in the ensemble, as a function of time, the number of atoms in the system, and as a function of the inter-pulse delay. We identify clear signatures of the enhanced cooperative dynamics for the pulse-driven inhomogeneous ensemble when dephasing in the dynamics of the ensemble is suppressed by a sufficient number of pulses during the individual emitter lifetime. These results demonstrate a post-fabrication tuning knob based on externally applied control fields for the dynamics of the many-emitter ensemble.  \\

\end{abstract}

\maketitle

%--------------------------------------------------------------------------
%--------------------------------------------------------------------------

\section{Introduction}
\label{sec:Introduction}

\noindent The dynamics of ensembles of quantum emitters and the emergence/destruction of cooperative behaviors such as that manifested through an early burst in the emitted radiation of a dense ensemble, superradiance, have been actively investigated for several decades~\cite{Dicke1954, HEPPLieb1973, GrossHaroche1982, BonifacioPRA1971_I, BonifacioPRA1971_II}. Interest in these fundamental effects has been greatly revived in recent years as a result of advances in device engineering and materials science\cite{MasonEtAlAsenjoGarciaPRL2020, douglasGreinerArXiv2026, Kim2025},  and as a result of novel promising applications in various areas ranging from cavity quantum electrodynamics~\cite{RaimondEtAlPRL1982, SlamaEtAlPRL2007,Delanty_2011, BaskicPRL2014} to quantum sensing~\cite{Colombo2022, Franke2023, Brinke2015, DeMille2024, KoppenhoeferPRL2023}. These cooperative properties often hinge on rather stringent conditions that can limit robust realizations to relatively narrow regions in parameter space and prevent widespread adoption/applications. Crucially, the emergence of cooperative dynamics in many emitter ensembles is often prevented by the disparity in the spatio-temporal configuration of individual emitters, such as variations arising from fluctuations in the surrounding bath. Indeed, the signature of cooperative dynamics will vanish as dephasing is boosted by the spatial separation of the emitters or as the differences between the local environments of individual emitters are increased~\cite{Sierra2022, MasonEtAlAsenjoGarciaPRL2020, Kim2025}.\\
External field control protocols have been widely used to protect a spin qubit, typically in the solid state, from surrounding magnetic moments that can induce deleterious decoherence effects~\cite{slichter1990principles, SlavaHansonScience2010, SlavaHansonScience2010, HansonDobrovitskiFeiguinGywatAwschalomScience2008, ViolaKnillLloydPRL1999}. Similarly, recent studies have demonstrated the capacity to mitigate spectral diffusion for a single quantum emitter or to tune its emission/absorption spectrum with pulse protocols and even to produce indistinguishable photons out of otherwise spectrally disparate/noisy quantum emitters~\cite{Fotso_TPI_PRB_2019, Fotso_noisyTLS2022, vuckovic2020}. Furthermore, in a dilute ensemble, it is natural then to expect that the emission/absorption spectrum can be readily refocused onto a shared spectral lineshape~\cite{FotsoEtal_PRL2016, unterguggenberger2026spectral, FotsoDobrovitski_Absorption, FotsoOtherPulsesJPhysB2018}. These advances bring about a rather intriguing prospect: Can we induce cooperative dynamics out of a dense inhomogeneous ensemble with an appropriate control protocol? If so, how is such a pulse-enhanced cooperative evolution manifested? Particularly given that the control field may lead to a systematic increase in the number of excitations in the system.\\
In the present paper, we consider an inhomogeneous ensemble of quantum emitters in the dense limit, where the spatial separation between emitters is much smaller than the typical emission wavelength, while the emission frequencies of individual emitters in the ensemble are spread over a frequency range extending over many linewidths. We assess the effect of an applied periodic pulse sequence. Using a combination of analytical and numerical methods, we track the dynamics of the system and examine the emitted radiation as a function of time as well as the properties of the many-emitter system. We find that for a relatively modest number of pulses per lifetime, the spectrally disparate ensemble can be made to behave as though it were homogeneous, exhibiting signatures of cooperative dynamics akin to that of a fully resonant dense system. These results demonstrate a novel tuning knob in the realization of coherent dynamics in many-emitter ensembles such as those of the solid state that are susceptible to extended spectral variability between individual emitters due to environmental fluctuations of the manufacturing imperfections\cite{NVCentersHansonGywatAwschalomPRB2006, NVCentersDOHERTY20131, NVCentersDolde2011, MaurerLukin_etAl_Science2012, QDotsLossDiVincenzoPRA1998, QDotsPetaGossard2005,QDotsYangDzurak2020}.\\
The remainder of the paper is organized as follows. In section~\ref{sec:ModelMethods}, we present the model for the many-emitter ensemble driven by a sequence of instantaneous $\pi-$pulses and discuss both of our numerical solutions: our semi-analytical solution based on a mean-field approach for the emitters degrees of freedom, and the quantum simulation that explicitly includes both the bosonic modes and the emitter degrees of freedom.  In section~\ref{sec:Results} we present our results for the pulse-driven ensemble and we end with our conclusions in section~\ref{sec:conclusion}.

\section{Model and Methods}
\label{sec:ModelMethods}

\noindent In the absence of any control protocol, we consider a system described by the many-atom extension of the Weisskopf-Wigner Hamiltonian~\cite{WeisskopfWigner1930, Dicke1954}:

\begin{eqnarray}
        H &=& \frac{\hbar}{2}\sum_i \omega_i\sigma_i^z +
        \hbar\sum_{\mathbf k \lambda}\omega_{\mathbf k} a_{\mathbf k \lambda}^\dagger a_{\mathbf k \lambda} \nonumber \\
        &-& i\hbar\sum_i \sum_{\mathbf k\lambda} g_{\mathbf k \lambda}^i(\sigma_i^+ + \sigma_i^-) \left( a^\dagger_{\mathbf k\lambda}e^{-i\mathbf k \cdot \mathbf r_i} - a_{\mathbf k\lambda}e^{i\mathbf k \cdot \mathbf r_i}\right). \nonumber \\
        & & 
    \label{eq:superradiant_hamiltonian}
\end{eqnarray}

\noindent Where $\sigma_i^z = |e\rangle\langle e|-|g\rangle\langle g|$ and $\sigma_i^+ = |e\rangle\langle g|$ ($\sigma_i^- = (\sigma_i^+ )^\dagger$) are the standard $z-$axis, and raising (lowering) operators for two-level atom $i$ at spatial location $r_i$ with emission frequency $\hbar\omega_i = E^i_e - E^i_g$. Here, $E^i_e$ and $E^i_g$ are respectively the excited and the ground state energy of atom $i$. $a_{\mathbf k\lambda}$ ($a_{\mathbf k\lambda}^\dagger$) is the annihilation (creation) operator acting on the photon mode with wavevector $\mathbf k$, frequency $\omega_{\mathbf k}$ and polarization $\lambda$ with coupling strength $g_{\mathbf k\lambda}^i$ to atom $i$.

\begin{figure}[t]
    \centering
    \includegraphics[width=\columnwidth]{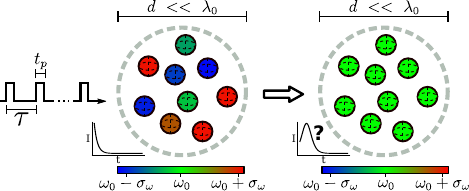}
    \caption{Schematic representation of a dense inhomogeneous ensemble of two-level systems where the emission frequencies of the individual emitters are distributed over a range $[\omega_0 - \sigma_{\omega}, \omega_0 + \sigma_{\omega}]$ and the characteristic spatial confinement of the emitters in the ensemble, $d$, is much less than the typical wavelength $\lambda_0$. We aim to assess the effect of a periodic pulse sequence on the inhomogeneous ensemble and the properties of the emitted radiation.}
    \label{fig:schematic}
\end{figure}

In what follows we set $\hbar = 1$, assume that all atoms and modes share a common axis of polarization, apply the long wavelength and rotating wave approximations, and enter the frame rotating at the frequency $\omega_0$ at which we will apply our control pulses. These assumptions amount to dropping the sums over $\lambda$, setting $e^{i\mathbf k\cdot\mathbf{r}_{ij}}\approx1$ for all $i$ and $j$, assuming the rotating wave approximation (RWA), and writing Hamiltonian~(\ref{eq:superradiant_hamiltonian}) as 
\begin{eqnarray}
        H &=& \frac{1}{2}\sum_i \Delta_i\sigma_i^z +
        \sum_k\Delta_{k} a_k^\dagger a_k \nonumber \\
        &-& \sum_i \sum_k g_k^i\left(\sigma_i^-a^\dagger_k - \sigma_i^+a_k\right)
    \label{eq:approximate_superradiant_hamiltonian}
\end{eqnarray}

\noindent where $ \Delta_{i,k} = \omega_{i,k} - \omega_0$ are respectively the detunings between atom $i$ or bosonic mode $k$ and the pulse carrier frequency. The coupling constants $g_k^i$ are related to the atomic lifetime $\Gamma_0^{-1}$ by Fermi's golden rule. In this work we use $\Gamma_0=1$ as our unit for energy, and inverse time. Thus, all results will be presented in terms of this unit. We have also confirmed that relaxing the RWA does not meaningfully affect the results in the present work (see Appendix~\ref{app:rwa}) . We are interested in the situation described schematically in Fig.~\ref{fig:schematic} where we illustrate the situation of a dense inhomogeneous ensemble of two-level systems where the emission frequencies of the individual emitters are distributed over a frequency range $[\omega_0 - \sigma_{\omega}, \omega_0 + \sigma_{\omega}]$. We aim to assess the effect of a periodic pulse sequence on the dynamics of the ensemble, to characterize the properties of the emitted radiation, and their dependence on the pulse sequence.
%\textcolor{blue}{cooperative dynamics of the ensemble and to characterize the properties of the emitted radiation}.
The pulse protocol is addressed by adding to the Hamiltonian~(\ref{eq:approximate_superradiant_hamiltonian}) the control field Hamiltonian: $H_c(t) = \Omega/2\sum_i[\sigma_i^++\sigma_i^-]$
where the Rabi frequency $\Omega(t)$ is such that $\Omega(t) = \pi/t_p$ over the pulse duration $\tau_p$ and $0$ otherwise. The results discussed here will focus on instantaneous $\pi$-pulses ($t_p\rightarrow0$) and we discuss finite width pulses in Appendix~\ref{app:finite}.

The emitted light intensity, or photon emission rate, is given by
\begin{equation}
    I(t) =  \Gamma_{0} \sum_{ij}\langle\sigma_i^+\sigma_j^-\rangle = I_{NC}(t) + I_C(t)
    \label{eqn:intensity}
\end{equation}
 \noindent where $I_{NC}(t)=\Gamma_0\sum_i\langle\sigma_i^+\sigma_i^-\rangle$ represents the non-coherent contribution to the intensity while $I_C(t)= \Gamma_{0} \sum_{i\neq j} \langle\sigma_i^+\sigma_j^-\rangle$ the coherent contribution, also referred to as the coherence of the system. $\Gamma_0$ appears in the definition of the coherence in place of $\Gamma_{ij}$ due to the dense ensemble limit.

We will track the dynamics of the system when it is initialized with all atoms in their excited states and all bosonic modes empty and allowed to evolve in time. To this end, we employ two complementary methods that we present briefly here with more extensive discussions provided in the appendices.

\subsection{Semi-Analytic Mean Field Approach}
\label{subsec:Analytics}

\noindent Using the Lindblad master equation, the time dependence of any $\mathcal O$ operator can be found using
\begin{equation}
   \langle \dot{\mathcal{O}}\rangle = i\langle{[H_{\text{dis}},\mathcal O]\rangle} + \Gamma_0\langle J_+\mathcal{O}J_- - \frac12\{J_+J_-,\mathcal{O}\}\rangle
    \label{eqn:master_eqn}
\end{equation}
%\noindent in which $\mathcal O$ is the operator for which we are calculating time evolution, 
\noindent $H_{\text{dis}} = \frac12\sum \Delta_i\sigma_i^z$ is the unitary evolution Hamiltonian that describes the dephasing via atomic detunings, and $J_\pm =\sum\sigma_i^\pm$ are the collective raising/lowering operators that encode superradiance in the limit of spatially co-located two-level emitters. 

Using a second order cumulant expansion, and defining,

\begin{align}
    I_{NC}^i = \langle\sigma_i^+\sigma_i^-\rangle &\text{  and  } I_C^{ij} = \langle\sigma_i^+\sigma_j^-\rangle
    \label{eqn:cij-and-nij}
\end{align}

we obtain the system of coupled differential equations:

\begin{eqnarray}
        \dot I_{C}^{ij} &=& i(\Delta_i-\Delta_j)I_{C}^{ij}+\frac{\Gamma_0}{2}\bigg\{(2I_{NC}^j-1)(I_{NC}^i + \sum_{\ell\ne i} I_{C}^{i\ell}) \nonumber \\  
        &+& (2I_{NC}^i-1)(I_{NC}^j+\sum_{k\ne j}I_{C}^{kj})\bigg\} \label{eq:cij_dot-and-nij_dot1} \\
        \dot I_{NC}^i &=& -\frac{\Gamma_0}{2}\bigg\{2I_{NC}^i+\sum_{\ell\ne i} I_{C}^{i\ell} + \sum_{k\ne i} I_{C}^{ki}\bigg\}.
    \label{eq:cij_dot-and-nij_dot2}
\end{eqnarray}

\noindent Note that Equations (\ref{eq:cij_dot-and-nij_dot1}) and  (\ref{eq:cij_dot-and-nij_dot2}) reduce to the well known system of equations characteristic of superradiant dynamics in the Dicke limit when $I_{NC}^i,I_{C}^{ij}\rightarrow I_{NC}, I_{C}$ and $\Delta_i-\Delta_j\rightarrow0$ for all $i$'s and $j$'s,
\begin{eqnarray}
%    \begin{align}
        \dot I_{C} &=& \Gamma_0(2I_{NC}-1)(I_{NC} + (N-1)I_{C})
        \label{eq:cij_dot-and-nij_dot-DICKE-limit1} \\
        \dot I_{NC} &=& -\Gamma_0(I_{NC}+(N-1)I_C).
%    \end{align}
    \label{eq:cij_dot-and-nij_dot-DICKE-limit2}
\end{eqnarray}

\noindent From the definitions (\ref{eqn:cij-and-nij}), we see that applying an $\pi$-pulse on the emitters has the effect: $I_{NC}^i\rightarrow 1- I_{NC}^i$ and $I_{C}^{ij}\rightarrow I_{C}^{ij*}$. 
Starting with initial conditions $I_{NC}^i(0)=1,~I_{C}^{ij}(0)=0$, we discretize the time axis into small timesteps and iteratively integrate Eqs.(\ref{eq:cij_dot-and-nij_dot1}) and  (\ref{eq:cij_dot-and-nij_dot2}) numerically between consecutive pulses and incorporate the above effects of the $\pi$ pulses at times $n\tau$, where $\tau$ is the inter-pulse delay  and $n = 1,2, ..., N_{p}$ with $N_{P}$ the number of pulses.

\begin{figure}[t] % [htbp]
    \centering
    \includegraphics[width=\columnwidth]{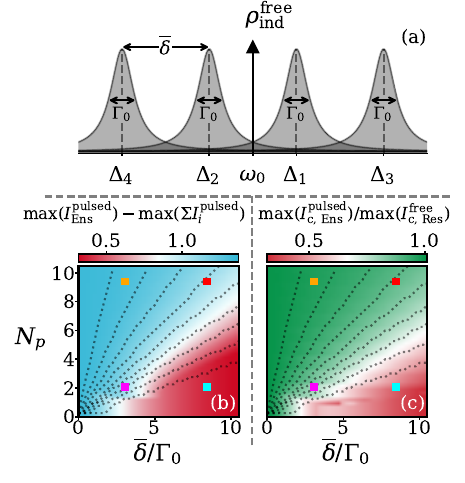}
    \caption{(a) We examine an ensemble of 4 emitters with different frequencies detuned from the pulse carrier frequency by $\Delta_i$ for emitter $i$, spread uniformly on the frequency axis with consecutive detunings separated by $\bar{\delta}$. The radiation intensity and coherence of this ensemble as a function of $N_p$, the number of pulses per lifetime, and the detuning spacing $\bar{\delta}$ is depicted in panels (b) and (c), respectively. 
The highlighted points correspond to detunings such that $\bar\delta=3$ with number of pulses per lifetime $N_p = \tau^{-1}= 2 $ (magenta), $~9.5$ (orange) and $\bar\delta=8.5$ with $N_p = 2$ (cyan) and $~9.5$ (red) and are further analyzed in Figs.~\ref{fig:intensity} and~\ref{fig:modes}.
  The dotted gray lines correspond to  different amounts of spectral overlap,  ranging from 10\% (lowest) to 90\% (highest), between the pulse-driven individual emitters that constitute the ensemble.} 
    \label{fig:heatmaps}
\end{figure}

In this way, we compute the intensity and its coherent and incoherent contributions for an ensemble of $N$ emitters with detuned emission frequencies distributed around the pulse carrier frequency as depicted in Fig.~\ref{fig:heatmaps}(a). The detuning of emitter $i$ from the pulse carrier frequency is $\Delta_i$, the detuning between emitters $i$ and $j$ is denoted $\delta_{ij}=\omega_i-\omega_j=\Delta_i-\Delta_j$. For an ensemble with a given number of emitters $N$, we use a constant separation between consecutive frequencies $\bar\delta$ as a measure of the frequency inhomogeneity in the ensemble.

%--------------------------------------------------------------------------
%--------------------------------------------------------------------------

\subsection{Digital Quantum Simulation}
\label{subsec:QuantumAlgo}

\noindent Extracting the full time dynamics of a cooperative $N$-emitter ensemble, regardless of spectral inhomogeneity, requires tracking a Hilbert space of size $2^N$ even after tracing out the radiative bath. This makes analytical solutions intractable for even modest system sizes and highlights the necessity of alternative, often numerical solutions~\cite{WODKIEWICZ1976574}. Furthermore, typical approximations performed in analytical solutions may prevent accurate system characterization in certain regimes e.g the Born-Markov approximation that breaks down for large $N$ systems~\cite{gross-haroche} or the mean field approximation which breaks down for small $N$ and/or long time. 

Here, we supplement our mean-field solution with a recently developed digital quantum simulation approach that tracks the state of the entire system, atoms and bosonic bath included~\cite{FotsoSPIE2025, yuvarajan2025, iglesiascardinale2026}. The method constructs a mapping of the bosonic modes onto qubit operators alongside the natural mapping of the emitters onto corresponding qubits. The time evolution is then carried out using a Trotter discretization of the time-evolution operator into small enough timesteps to ensure negligible discretization errors. The effect of an instantaneous $\pi$-pulse at time $n \tau$ is equivalent to applying an $X$ ($NOT$) gate on all qubits modeling two level emitters between the Trotter step corresponding to time $n\tau$ and that to $n\tau+\Delta t$. While the  method requires the discretization of the frequency axis and is limited to a small number of available bosonic modes or to a truncated occupation number for each mode, as well as to a finite discretization step, $\Delta t$, of the time axis, its limitations can be controlled for a small number of emitters~\cite{iglesiascardinale2026}, delivering reliable results that complement the mean-field solution.

\section{Results}
\label{sec:Results}

%Coherence
\noindent The coherence $I_C(t)$ should remain close to zero for an incoherent ensemble, where the coupling between the emitters through the bosonic bath is negligible and they remain independent, as is the case when the spectral overlap between emitters in the ensemble vanishes. For the dense resonant ensemble of emitters without any additional driving protocol, cooperative dynamics of the ensemble is manifested through an enhancement of the coherence as well as an intensity burst with a peak maximum that scales like $N^2$ with $N$ the number of emitters participating in the cooperative dynamics.

\begin{figure}[t]
    \centering
    \includegraphics[width=\columnwidth]{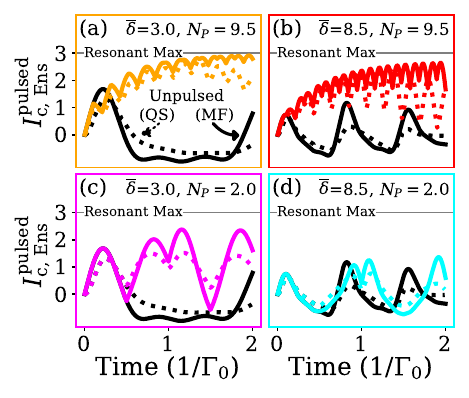}
    \caption{Coherence as a function of time of the pulse driven ensemble for the points highlighted in Fig.~\ref{fig:heatmaps}. Each panel compares the pulse driven system to the unpulsed or freely evolving system starting from the same initial configuration. Solid lines correspond to the mean-field solution while the dotted lines correspond to the quantum simulation solution. The solid gray horizontal line indicates the maximum of the coherence for a fully resonant non-driven system. Clearly, for short inter-pulse delays, the maximum coherence is approached in the driven ensemble. As the detuning parameter $\bar\delta$ is increased, shorter inter-pulse delays are required to achieve comparable enhancements.}
    \label{fig:intensity}
\end{figure}

\begin{figure}[htbp]
    \centering
    \includegraphics[width=\columnwidth]{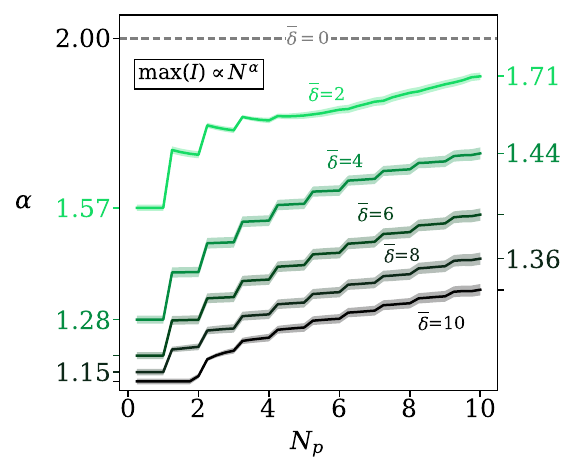}
    \caption{Exponent $\alpha$ for the scaling of the maximum in the intensity, of a spectrally inhomogeneous ensemble as a function of the number of emitters: $f_{\text{fit}}=A(N+1)^\alpha$. 
    Applying a sequence of periodic $\pi$-pulses to the ensemble enhances the scaling of the detuned ensembles as the inter-pulse delay decreases.}
    \label{fig:scaling}
\end{figure}

\begin{figure}[htbp]
    \centering
    \includegraphics[width=\columnwidth]{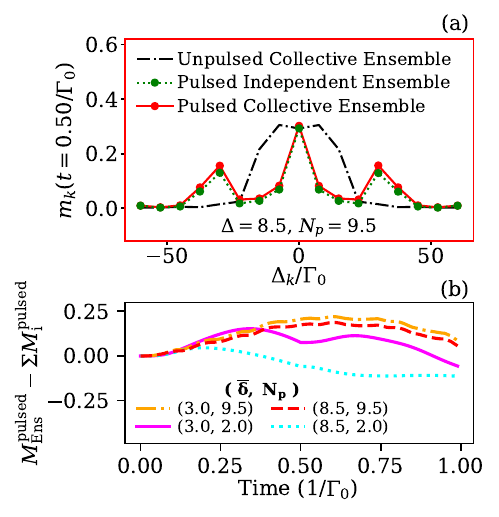}
    \caption{Mode occupation of the radiation bath in the quantum simulation as a function of time for $\bar\delta=3.0$ and $\tau=1/9.5$. Collective effects result in a greater number of photons present in the system at early times for the pulsed ensemble than the sum of pulsed individuals. This is shown in panel (a) which highlights this phenomenon at the points of interest described in prior figures. Panel (b) depicts the occupation of each bosonic mode for the pulsed and unpulsed ensemble as well as the sum of individual pulsed emitters. This corresponds to the expected emission spectrum at time $t=0.5/\Gamma_0$.}
    \label{fig:modes}
\end{figure}

In Fig.~\ref{fig:heatmaps}, we examine an ensemble of 4 emitters with different frequencies detuned from the pulse carrier frequency by $\Delta_i$ for emitter $i$, spread uniformly on the frequency axis with consecutive detunings separated by $\bar{\delta}$. The radiation intensity and coherence of this ensemble as a function of $N_p$, the number of pulses per lifetime, and the detuning spacing $\bar{\delta}$ are depicted in panels (b) and (c), respectively.
The highlighted points correspond to detunings such that $\bar\delta=3$ with number of pulses per lifetime $N_p = \tau^{-1}= 2 $ (magenta), $~9.5$ (orange) and $\bar\delta=8.5$ with $N_p = 2$ (cyan) and $~9.5$ (red) and are further analyzed in Figs.~\ref{fig:intensity} and~\ref{fig:modes}.
d by numerically integrating the overlap of the pulsed independent emission spectra of all atoms in the ensemble as derived in Ref.~\cite{FotsoDobrovitski_Absorption}. These results, along with those of Fig.~\ref{fig:intensity} and Fig.~\ref{fig:scaling} are obtained with the mean-field solution, and so are easily extended to a large number of emitters (see Appendix~\ref{app:more_heatmaps}). Note, however, that for a small number of emitters we get very good agreement with the quantum simulation even though the latter addresses a finite size system, while the former is formulated in the thermodynamic limit.

To further characterize the effects of the pulse sequence on the system depicted in Fig.~\ref{fig:heatmaps}-(a), we examine the coherence as a function of time at the points highlighted in Fig.~\ref{fig:heatmaps} - (b) and (c). 
This is shown in Fig.~\ref{fig:intensity}. Each panel compares the pulse driven system to the unpulsed or freely evolving system starting from the same initial configuration. Solid lines correspond to the mean-field solution, while the dotted lines correspond to the quantum simulation solution. The solid gray horizontal line indicates the maximum of the coherence for a fully resonant non-driven system. Clearly, for short inter-pulse delays, the maximum coherence is approached in the driven ensemble. As the detuning parameter $\bar\delta$ is increased, shorter inter-pulse delays are required to achieve comparable enhancements.

Taking the difference between an observable of the pulsed ensemble and the sum of the same observable from the individual pulsed but independent constituents of the ensemble can also effectively elucidate the enhancement of cooperative dynamics on that particular observable. As expected, for the intensity, this difference leaves only the coherent contribution and thus we observe that the enhancements observed in the coherence are also manifested in a similar manner in the overall emission intensity as a function of time.

The maximum of the emission intensity for an ensemble of $N$-emitters scales roughly as~\cite{breuer-petruccione}
\begin{equation}
    \max(I)\propto N^\alpha.
    \label{eqn:scaling_fit}
\end{equation}
\noindent Where for a superradiant ensemble in the Dicke limit $\alpha=2$, and for an ensemble of independent (non-coherent) emitters $\alpha=1$. The inhomogeneous spectral distribution  of the system described above in Fig.~\ref{fig:heatmaps} (a) will lead to a gradual change from the quadratic scaling of the interacting ensemble as $\bar{\delta}$ increases due to dephasing in the dynamics of the emitters in the ensemble. In Fig.~\ref{fig:scaling}, we show how this scaling is modified for different values of $\bar{\delta}$ as the number of pulses per lifetime is varied. We observe that for modest $\bar{\delta}$ values e.g for $\bar{\delta} = 2$, we start from a situation where the $\alpha \to~ \sim1.6$ and,  even at low $N_p$ values, we already notice a dramatic increase in the exponent toward $\alpha \sim 2$ as $N_p$ is increased. This trend persists as $\bar{\delta}$ increases although starting from a lower $\alpha$ value. This is highlighted in Fig.~\ref{fig:scaling}, where the integration of equations (\ref{eq:cij_dot-and-nij_dot1}) and (\ref{eq:cij_dot-and-nij_dot2}) is carried out for ensembles of size $N=2$ to $N=25$ and the scaling exponent $\alpha$ is extracted by a standard fitting procedure with the function $f_{\text{fit}}=A(N+1)^\alpha$. Note that our system is defined with the linewidth $\Gamma_0$ as our unit of frequency. Thus, for the distribution of atoms in frequency space studied here, any detuning $\bar\delta>2$ is significant (this is evidenced in panels (b) and (c) of Fig.~\ref{fig:heatmaps}).

Using the digital quantum simulation approach, we are able to track the dynamics of the entire \textit{"emitters + bosonic modes"} system. A key advantage of this method is that it provides explicit access to the dynamics of the bosonic radiation modes. The total number of photons present in the system at time $t$ is given by:
\begin{equation}
    M(t) =\sum_k \langle a_k^\dagger a_k\rangle = \sum_k m_k
    \label{eqn:photon_count}
\end{equation}
\noindent where $m_k(t)$ represents the photon occupation of mode $k$ at time $t$. 

We study a four-atom system similar to that of Fig.~\ref{fig:heatmaps}(a) with the quantum algorithm. The four atoms are coupled to a discretized bath of fifteen bosonic modes. The central mode ($m_0$) is placed at the pulse carrier frequency and the second-to-outermost modes ($m_{-6}$ and $m_{6}$, respectively) are placed at the second order satellite peaks ($\omega_\text{sat}=\pm n\pi/\tau$ with $n=2$). The location of all other modes are determined by setting the inter-mode frequency difference to a constant consistent with the definition of $m_{-6}$ and $m_{6}$.

An ensemble of $N$ emitters coupled to a bosonic bath will emit $N$ photons regardless of whether cooperative behavior is present in the system. Cooperative behavior affects only the rate at which these photons are emitted, with strong inter-emitter coherence leading to faster photon emission at early times. We establish this behavior in the pulse-driven ensemble %this effect arises due to the pulse sequence applied to a diffuse ensemble due 
by taking the difference between the total mode occupation of the pulsed-driven ensemble $M_{\text{Ens}}^{Pulsed}$ and the total mode occupation of the pulsed independent ensemble $\sum_i M_i^{Pulsed}$. This quantity is shown as a function of time in Fig.~\ref{fig:modes} for $\bar\delta=3.0$ and $\tau=1/9.5$ (orange), $\bar\delta=8.5$ and $\tau=1/9.5$ (red), $\bar\delta=3.0$ and $\tau=1/2.0$ (magenta), $\bar\delta=8.5$ and $\tau=1/2.0$ (cyan). We can see that the parameters with more enhanced coherence also display a faster increase in the mode occupation.

\section{Conclusion}
\label{sec:conclusion}

\noindent We have demonstrated the enhancement of cooperative dynamics in a dense spectrally inhomogeneous ensemble when it is driven by a periodic sequence of $\pi$-pulses. Using both a semi-analytical mean-field solution and a digital quantum simulation, we analyzed the coherence of the pulse-driven ensemble and the intensity of the emitted radiation as a function of the disparity in the ensemble, as a function of time, the number of atoms in the system, and as a function of the inter-pulse delay. Through suitable comparisons with the driven individual emitters and with a non-driven resonant ensemble, we identify clear signatures of the enhanced cooperative dynamics for the pulse-driven system when dephasing in the dynamics of the ensemble is suppressed by a sufficient number of pulses during the individual emitter lifetime.
These results demonstrate a novel tuning knob in the realization of coherent dynamics in many-emitter ensembles such as those of the solid state that are susceptible to extended spectral variability between individual emitters due to environmental fluctuations or to manufacturing imperfections\cite{NVCentersHansonGywatAwschalomPRB2006, NVCentersDOHERTY20131, NVCentersDolde2011, MaurerLukin_etAl_Science2012, QDotsLossDiVincenzoPRA1998, QDotsPetaGossard2005,QDotsYangDzurak2020}

%--------------------------------------------------------------------------
%--------------------------------------------------------------------------

\section{Acknowledgments}
\noindent This work was supported by the National Science Foundation under Grants No. PHY-2014023 and No. QIS-2328752. The authors thank A. Asenjo-Garcia, J. Marino, D. Schneble and S. Gallis for helpful discussions. \\

%--------------------------------------------------------------------------
%--------------------------------------------------------------------------
%\bibliographystyle{apsrev4-2}
\bibliography{mainBib}

%--------------------------------------------------------------------------
%--------------------------------------------------------------------------
%--------------------------------------------------------------------------
%--------------------------------------------------------------------------
\newpage
\onecolumngrid
\appendix

\newpage
\section{Derivation of Eqns.~\ref{eq:cij_dot-and-nij_dot1} and ~\ref{eq:cij_dot-and-nij_dot2}}
\label{app:mean_field_derivation}

The hallmark characteristic defining superradiant emission is the dominance of inter-emitter coherence $I_C(t)$ over pure atomic population $I_{NC}$ in the emission intensity,

\begin{equation}
    I(t) = \sum_{ij}\Gamma_{ij}\langle\sigma_i^+\sigma_j^-\rangle = \underbrace{\sum_{i}\Gamma_{ii}\langle\sigma_i^+\sigma_i^-\rangle}_{I_{NC}} + \underbrace{\sum_{i\ne j}\Gamma_{ij}\langle\sigma_i^+\sigma_j^-\rangle}_{I_C(t)} = \Gamma_0\left\{\sum_iI_{NC}^i+\sum_{ij}I_C^{ij} \right\}
    \label{eqn:app1_intensity}
\end{equation}

\noindent where in the last equality we have assumed that the only decay channel present is $\Gamma_{ij}=\Gamma_0$, or the superradiant decay channel. This is valid when considering a spatially dense ensemble as is depicted in Fig.~\ref{fig:schematic} and discussed as the primary system the main text. The effects of spatial distribution, though not directly pertinent to the spectral dephasing problem addressed in the main text, have been studied at length in the literature~\cite{Masson2022}, and the methods used to study that system inspire much of the work we do here.

In particular, we implement the master equation approach to derive the time dynamics of the intensity. Mathematically, for an arbitrary operator $\mathcal O$, this amounts to solving

\begin{equation}
   \langle \dot{\mathcal{O}}\rangle = i\langle{[H_{\text{dis}},\mathcal O]\rangle} + \Gamma_0\langle J_+\mathcal{O}J_- - \frac12\{J_+J_-,\mathcal{O}\}\rangle,
    \label{eqn:app1_master_eqn}
\end{equation}

\noindent which is the well known Lindblad master equation with $H_{\text{dis}} = -\frac12\sum_i \Delta_i\sigma_i^z$ the unitary evolution Hamiltonian that describes atomic dephasing and $J_\pm =\sum\sigma_i^\pm$ the collective raising/lowering operators that encode superradiance. To find the time evolution of the intensity we set $\mathcal O = \sigma_i^+\sigma_j^-$ and find

\begin{align}
    \frac{d}{dt}\langle\sigma_i^+\sigma_j^-\rangle &= \frac{i}{2}(\delta_i-\delta_j)\langle \sigma_i^+\sigma_j^-\rangle + \Gamma_0\sum_{k\ell} \underbrace{\left(\langle\sigma_k^+\sigma_i^+\sigma_j^-\sigma_\ell^-\rangle -\frac{1}{2}\langle\sigma_k^+\sigma_\ell^-\sigma_i^+\sigma_j^-\rangle -\frac{1}{2}\langle\sigma_i^+\sigma_j^-\sigma_k^+\sigma_\ell^-\rangle\right)}_{A_{ij}}
    \label{eqn:app1_cij_dot_general}
\end{align}

\noindent in which four-operator expectation values have arrived in the sum. We name the collection of these operators $A_{ij}$ for ease of reference. This is an apparent manifestation of the BBGKY heirarchy problem, and we must handle the problem differently in the two cases in which $i=j$ and $i\ne j$. When solving for $\dot I_{NC}^i(t)$ (or equivalently when $i=j$ in Eqn.~\ref{eqn:app1_cij_dot_general}), we use the facts that $\sigma_i^+\sigma_i^- = \frac{1}{2}(1+\sigma_i^z)$ and $[\sigma_i^z,\sigma_j^\pm]=\pm2\delta_{ij}\sigma_i^\pm$ to nicely simplify $A_{ij}=A_{ii}$ to

\begin{equation}
    \begin{aligned}
        A_{ii} &=\frac{1}{2}\langle\sigma_k^+\sigma_i^z\sigma_\ell^-\rangle -\frac{1}{4}\langle\sigma_k^+\sigma_\ell^-\sigma_i^z\rangle -\frac{1}{4}\langle\sigma_i^z\sigma_k^+\sigma_\ell^-\rangle\\
        &= \frac{1}{2}(1-\delta_{ik})(1-\delta_{i\ell})\langle\sigma_k^+\sigma_i^z\sigma_\ell^-\rangle -\frac{1}{2}\delta_{ik}\langle\sigma_i^+\sigma_\ell^-\rangle-\frac{1}{2}\delta_{i\ell}\langle\sigma_k^+\sigma_i^-\rangle\\
        &~ -\frac{1}{4}\langle\sigma_k^+\sigma_i^-\rangle\delta_{i\ell} - \frac{1}{4}(1-\delta_{i\ell})\langle\sigma_k^+\sigma_i^z\sigma_\ell^-\rangle\\
        &~-\frac{1}{4}\langle\sigma_i^+\sigma_\ell^-\rangle\delta_{ik} - \frac{1}{4}(1-\delta_{ik})\langle\sigma_k^+\sigma_i^z\sigma_\ell^-\rangle\\
        &= -\frac{1}{2}(\delta_{ik}\langle \sigma_i^+\sigma_\ell^-\rangle + \delta_{i\ell}\langle\sigma_k^+\sigma_i^-\rangle).
    \end{aligned}
    \label{eqn:app1_Omega,i=j}
\end{equation}

For the case of $i\ne j$, or when solving for $\dot I_{C}^{ij}(t)$, we must utilize a cumulant expansion,

\begin{equation}
    \langle\sigma_i\sigma_j\sigma_k\sigma_\ell\rangle \approx \langle\sigma_i\sigma_j\rangle\langle\sigma_k\sigma_\ell\rangle + \langle\sigma_i\sigma_k\rangle\langle\sigma_j\sigma_\ell\rangle +  \langle\sigma_i\sigma_\ell\rangle\langle\sigma_j\sigma_k\rangle,
    \label{eqn:app1_meanfield}
\end{equation}

\noindent to achieve closure. This approximation given by Eqn.~\ref{eqn:app1_meanfield} is valid only at early times (as $t\rightarrow0$), but this is the exact regime in which we're interested~\cite{Sanchez_Barquilla_2020, Kira_Koch_2011, Kira_Koch_2008}. With this we can solve for $A_{i\ne j}$.

\begin{equation}
    \begin{aligned}
        A_{ij}&\approx\langle\sigma_k^+\sigma_j^-\rangle\langle\sigma_i^+\sigma_\ell^-\rangle-\frac{1}{2}\langle\sigma_k^+\sigma_j^-\rangle\langle\sigma_\ell^-\sigma_i^+\rangle-\frac{1}{2}\langle\sigma_j^-\sigma_k^+\rangle\langle\sigma_i^+\sigma_\ell^-\rangle\\
         &= \langle\sigma_k^+\sigma_j^-\rangle\langle\sigma_i^+\sigma_\ell^-\rangle \\
         &~- \frac{\delta_{i\ell}}{2}\langle\sigma_k^+\sigma_j^-\rangle+\frac{\delta_{i\ell}}{2}I_{NC}^i\langle\sigma_k^+\sigma_j^-\rangle-\frac{1}{2}(1-\delta_{i\ell})\langle\sigma_k^+\sigma_j^-\rangle\langle\sigma_i^+\sigma_\ell^-\rangle\\
         &~- \frac{\delta_{jk}}{2}\langle\sigma_i^+\sigma_\ell^-\rangle+\frac{\delta_{jk}}{2}I_{NC}^j\langle\sigma_i^+\sigma_\ell^-\rangle-\frac{1}{2}(1-\delta_{jk})\langle\sigma_k^+\sigma_j^-\rangle\langle\sigma_i^+\sigma_\ell^-\rangle\\
         &= \langle\sigma_k^+\sigma_j^-\rangle\langle\sigma_i^+\sigma_\ell^-\rangle -\frac{\delta_{i\ell}}{2}(1-I_{NC}^i)\langle\sigma_k^+\sigma_j^-\rangle  -\frac{\delta_{jk}}{2}(1-I_{NC}^j)\langle\sigma_i^+\sigma_\ell^-\rangle\\
         &~-\frac{1}{2}(2-\delta_{i\ell} - \delta_{jk})\langle\sigma_k^+\sigma_j^-\rangle\langle\sigma_i^+\sigma_\ell^-\rangle\\
         &=\frac{1}{2}(\delta_{i\ell} + \delta_{jk})\langle\sigma_k^+\sigma_j^-\rangle\langle\sigma_i^+\sigma_\ell^-\rangle -\frac{\delta_{i\ell}}{2}(1-I_{NC}^i)\langle\sigma_k^+\sigma_j^-\rangle  -\frac{\delta_{jk}}{2}(1-I_{NC}^j)\langle\sigma_i^+\sigma_\ell^-\rangle\\
    \end{aligned}
    \label{eqn:app1_Omega,i!=j}
\end{equation}

At this point we can rewrite Eqn.~\ref{eqn:app1_cij_dot_general} for the case where $i\ne j$ (the coherences)

\begin{equation}
    \begin{aligned}
        \dot I_{C}^{ij} = i(\Delta_i-\Delta_j)I_{C}^{ij} + \frac{\Gamma_0}{2}\sum_{k\ell}\bigg\{(\delta_{i\ell} + \delta_{jk})I_{C}^{kj}I_{C}^{i\ell} -\delta_{i\ell}(1-I_{NC}^i)I_{C}^{kj}  - \delta_{jk}(1-I_{NC}^j)I_{C}^{i\ell}\bigg\}
    \end{aligned}
    \label{eqn:app1_cij_dot}
\end{equation}

\noindent and the case $i=j$ (the atomic level populations)

\begin{equation}
    \begin{aligned}
        \dot I_{NC}^i &=-\frac{\Gamma_0}{2}\sum_{k\ell}\{\delta_{ik} I_{C}^{i\ell} + \delta_{i\ell}I_{C}^{ki}\}.
    \end{aligned}
    \label{eqn:app1_ni_dot}
\end{equation}

We simplify Eqns. \ref{eqn:app1_cij_dot} and \ref{eqn:app1_ni_dot} by noting

\begin{equation}
    \begin{aligned}
        I_{C}^{ij} = (I_{C}^{ji})^*,~\text{and}
    \end{aligned}
    \label{eqn:app1_simplyifying-fact-1}
\end{equation}

\begin{equation}
    \sum_j I_{C}^{ij} =I_{NC}^i + \sum_{j\ne i}I_{C}^{ij}
    \label{eqn:app1_simplyifying-fact-2}
\end{equation}

\noindent with which we can finally write

\begin{subequations}
    \begin{align}
        \dot I_{C}^{ij} &= i(\Delta_i-\Delta_j)I_{C}^{ij}+\frac{\Gamma_0}{2}\left\{(2I_{NC}^j-1)(I_{NC}^i + \sum_{\ell\ne i} I_{C}^{i\ell}) + (2I_{NC}^i-1)(I_{NC}^j+\sum_{k\ne j}I_{C}^{kj})\right\}\\
        \dot I_{NC}^i &= -\frac{\Gamma_0}{2}\left\{2I_{NC}^i+\sum_{\ell\ne i} I_{C}^{i\ell} + \sum_{k\ne i} I_{C}^{ki}\right\}
    \end{align}
    \label{eqn:app1_cij_dot-and-nij_dot}
\end{subequations}

In the limiting case of non-interacting atoms, then $I_{C}^{ij}=0$ at all times and $\dot I_{NC}^i = -\Gamma_0I_{NC}^i$ which implies $I_{NC}^i(t)=e^{-\Gamma_0 t}$. This is exactly the expected exponential decay, and the sum over all emitters would simply yield $N$ times this function, which is again exactly what s to be expected.

It is clarifying now to take the Dicke limit in which the atoms become indistinguishable. This means $I_{NC}^i\rightarrow I_{NC}$, $I_{C}^{ij}\rightarrow I_{C}$ and $\Delta_i-\Delta_j \rightarrow 0$. In this limit

\begin{subequations}
    \begin{align}
        \dot I_{C} &= \Gamma_0(2I_{NC}-1)(I_{NC} + (N-1)I_{C})\\
        \dot I_{NC} &= -\Gamma_0(I_{NC}+(N-1)I_{C})
    \end{align}
    \label{eqn:app1_cij_dot-and-nij_dot-DICKE-limit}
\end{subequations}
%--------------------------------------------------------------------------
%--------------------------------------------------------------------------
\newpage
\section{Dependence on the Number of Emitters}
\label{app:more_heatmaps}

A small system of four spectrally diffuse emitters is studied in depth in the main text. Larger system sizes, which are of significant interest experimentally, are difficult to capture theoretically. The mean field approach detailed in the main text and Appendix~\ref{app:mean_field_derivation} is able to capture the dynamics of modestly larger systems while remaining computationally viable. 

\begin{figure}[h]
    \centering
    \includegraphics[width=6.5in]{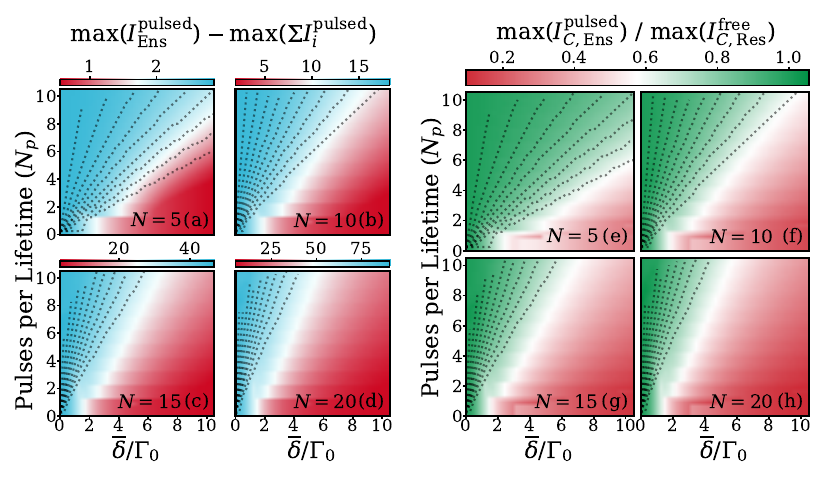} %manyheatmaps_combo.pdf}
    \caption{An ensemble of $N$ spectrally diffuse emitters with inter-emitter detuning $\bar\delta$ are prepared in the excited state and pulsed $N_P$ times per independent lifetime. The sum of peak emission intensities of independent, detuned emitters is subtracted from the peak intensity of the ensemble with identical atomic detunings and is presented in panels (a)-(d). The peak coherence of the diffuse ensemble, normalized by the peak of the unpulsed, resonant ensemble, is presented in panels (e)-(h). The dotted gray lines in all panels represent the effective overlap of emission spectra of independent, pulsed emitters. The lowest line represents 10\% total overlap and the highest represents 90\%.}
    \label{fig:whole_lotta_heatmaps}
\end{figure}

As the system size grows, the pulse rate necessary for an enhancement in coherence increases. This is due to the chosen distribution of emitters in frequency space. For a system with $N$ emitters, the greatest detuning from the pulse carrier frequency is on the order of $\Delta_{N}\sim N\bar\delta/2$ (see Fig.~\ref{fig:heatmaps}(a)). This result is made evident in Fig.~\ref{fig:whole_lotta_heatmaps} in which the same inter-emitter detuning and pulse rate parameter ranges as those in the main text are scanned for ensembles of sizes $N=5$, $10$, $15$, and $20$. The enhancement in the intensity is shown in panels (a)-(d), and that for the coherence in panels (e)-(h). By numerically integrating the overlap of the emission spectra of independent, detuned atoms, we see that an induced overlap of only 10\% provides a significant enhancement regardless of system size.

%--------------------------------------------------------------------------
%--------------------------------------------------------------------------
\newpage
\section{Remarks on the agreement (and disagreement) between the Mean Field and Quantum Simulation approaches}
\label{app:MFQA_comparison}

Mean field based solutions of many-body quantum systems as used in the main text are known to breakdown at late times and for small system sizes. While this is not such a concern when finding the maximum intensity of a superradiant ensemble (which occurs at early times), it partially explains the marginal deviation seen between the methods in Fig.~\ref{fig:intensity} of the main text. The remainder of the deviation is accounted for by the field/mode truncation in the quantum simulation. The limited number of modes modeling the vacuum in this case causes atomic revival at late times, and the limited range of mode frequencies induces a slight delay in decay. Quantitative and qualitative agreement between the methods despite these complications is therefore strong validation of the results presented in the main text. 

\begin{figure}[h]
    \centering
    \includegraphics[width=5in]{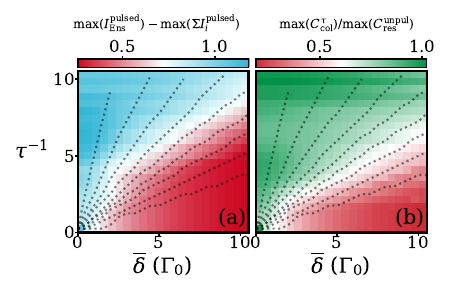} %QAheatmaps.pdf}
    \caption{(a) Sum of maximum intensity achieved by individual, detuned emitters subtracted from the pulsed, diffuse ensemble to extract cooperative dynamics. (b) Maximum coherence achieved by diffuse ensemble suffering periodic $\pi$-pulse sequence normalized by the maximum coherence achieved by the undriven, resonant system.}
    \label{fig:QAheatmaps}
\end{figure}

Fig.~\ref{fig:heatmaps} of the main text characterizes the maximum intensity achieved by a spectrally inhomogeneous, pulsed, four-atom ensemble with a fairly dense scan over a relatively large parameter space. This is obtained by integrating the system of equations defined in Eqns.~\ref{eq:cij_dot-and-nij_dot1} and ~\ref{eq:cij_dot-and-nij_dot2}. To validate the behavior shown there via the quantum simulation, we present a coarser scan over the same parameter range in Fig.~\ref{fig:QAheatmaps}. 

There is remarkable qualitative agreement between panels (a) and (b) of Fig.~\ref{fig:QAheatmaps}. In both cases, sequential application of $\pi$-pulses is shown to enhance the maximum coherence achieved by the ensemble, with resonant conditions being approached for fast enough pulse sequences. The quantitative differences presented can be attributed to the different approximations made between the methods. 

%--------------------------------------------------------------------------
%--------------------------------------------------------------------------
\newpage
\section{Finite Width Pulses}
\label{app:finite}

The approximation of instantaneous $\pi$-pulses utilized throughout the manuscript can be lifted within the digital quantum simulation. The familiar ensemble of four atoms coupled to a bath of fifteen bosonic modes studied in the main text is again studied here, however we apply finite width $\pi$-pulses in place instantaneous ones. We find that lifting this approximation does not significantly impact the enhancement of coherence in the ensemble when the pulse width is less than a quarter of the inter-pulse delay.

\begin{figure}[h]
    \centering
    \includegraphics[width=6.5in]{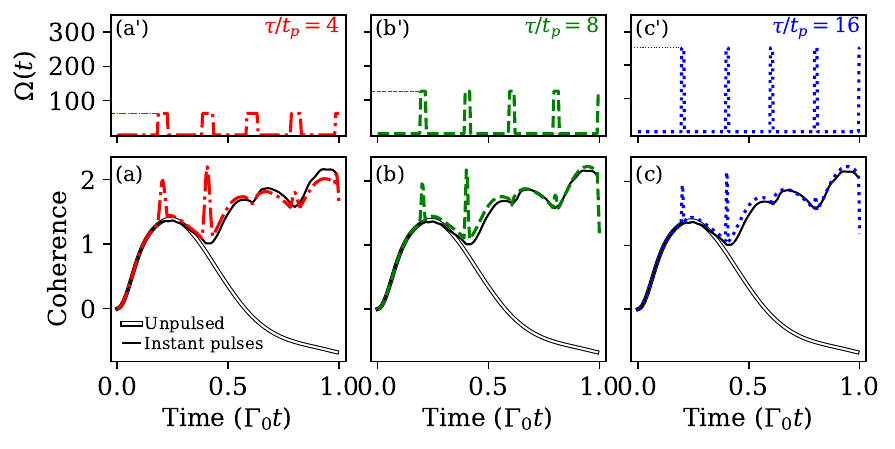} %finite_pulse.pdf}
    \caption{Panels (a), (b), and (c) present the coherence achieved by the unpulsed ensemble (solid white), the ensemble suffering a sequence of instantaneous $\pi$-pulses with interpulse delay $\tau=1/(5\Gamma_0)$ (solid black), and the ensemble suffering a sequence finite width $\pi$-pulses of the same interpulse delay and delay-to-width ratios of $\tau/t_p=4$ (dash-dotted red), $\tau/t_p=8$ (dashed green), and $\tau/t_p=16$ (dotted blue). Panels (a$'$), (b$'$), and (c$'$) show the driving strength $\Omega(t)$ for each of the corresponding finite-width drives.}
    \label{fig:variable_finite_pulse}
\end{figure}

Finite width $\pi$-pulses are added simply by implementing a time dependent drive in Hamiltonian~\ref{eq:superradiant_hamiltonian},

\begin{equation}
    H_d = \frac{\Omega(t)}{2} \sum_i\sigma_i^x
    \label{eqn:drive}
\end{equation}

\noindent where $\Omega(t)=\pi/t_p$ during the pulse and $0$ otherwise. Three applications of this time dependent drive are presented in Fig.~\ref{fig:variable_finite_pulse}. There, we see that a interpulse-delay-to-pulse-width ratio $\tau/t_p=4$ nearly identically replicates the effect of instantaneous pulses with alignment only improving as the ratio increases (pulse width decreases).

%--------------------------------------------------------------------------
%--------------------------------------------------------------------------
\newpage
\section{Average Hamiltonian}
\label{app:avg}

The arguments presented in the main text rely on dynamical rephasing of spectrally inhomogeneous ensembles via a control pulse protocol. It is insightful, then, to see that the net dephasing effects which arise from the atomic term of the bare Hamiltonian vanish at lowest order in the time averaged Hamiltonian after a period of twice the inter-pulse delay. The higher order corrections scale with increasing powers of $\tau$, and thus the approximation becomes stronger as the pulse rate is increased. To show this, we utilize the Magnus expansion in which the average Hamiltonian after a full $2\tau$ pulse sequence (first pulse applied at $t=\tau$, second at $t=2\tau$) is given by~\cite{Magnus1954, Viola1999, Brinkmann2016}

\begin{equation}
    \begin{aligned}
        \bar H &= H^{(0)}+ H^{(1)} + H^{(2)} + \cdots \\
        &=\left(\frac{1}{2\tau}\int_0^{2\tau}dtH(t)\right) + \left(\frac{1}{4i\tau}\int_0^{2\tau}dt\int_0^tdt'[H(t),H(t')]\right) \\
        &\quad+ \left(-\frac{1}{12\tau}\int_0^{2\tau}dt\int_0^tdt'\int_0^{t'}dt''([H(t),[H(t'),H(t'')]]+[[H(t),H(t')],H(t'')])\right) + \cdots
    \end{aligned}
    \label{eqn:magnus}
\end{equation}

The Hamiltonian including the time-dependent pulse protocol is given by

\begin{equation}
    H = \frac12\sum_i\omega_i\sigma_i^z + \sum_k\omega_ka_k^\dagger a_k -i\sum_{ik}g_k(\sigma_++\sigma_-)(a_k^\dagger-a_k) + H_p(t)
    \label{eqn:pulsed-hamiltonian}
\end{equation}

\noindent in the lab frame. $H_p(t)=\Omega(t)\sum\sigma_i^x = (\pi/t_p)\cos(\omega_dt)\sum\sigma_i^x$ during the pulse duration $t_p$ and $H_p(t)=0$ otherwise is the term responsible for applying a $\pi$-pulse at every time $n\tau$ with $n\in\mathbb{Z}^+$. Entering the frame rotating at the driving frequency $\omega_d$ with the transformation $U_d=\exp(i\omega_dt[\frac12\sum_i\sigma_i^z + \sum_ka_k^\dagger a_k])$, the toggling frame with transformation $U_p = \exp(it\Omega(t)\sum\sigma_i^x/2)$ (in the limit $t_p\rightarrow0$ and $t\Omega(t)\rightarrow\pi$ for a $\pi$-pulse protocol), and applying the rotating wave approximation, we decompose the full Hamiltonian into its time-independent forms before and after the first pulse is applied at time $t=\tau$. These are deemed $H_{\text{even}}$ and $H_{\text{odd}}$, respectively, as they are also the forms of the Hamiltonian after an arbitrary even/odd number of pulses.

\begin{equation}
    \begin{aligned}
         H_{\text{even}} &= \underbrace{\frac{1}{2}\sum_i \Delta_i\sigma_i^z}_{H_{\text{A}}} + \underbrace{\sum_k \Delta_k a_k^{\dagger} a_k}_{H_{\text{B}}} \underbrace{-i \sum_{ik} g_k (\sigma_i^-a_k^\dagger - \sigma_i^+a_k)}_{H_{\text{I}}^{\text{even}}}\\
         H_{\text{odd}} &= -\underbrace{\frac{1}{2}\sum_i \Delta_i\sigma_i^z}_{H_{\text{A}}} + \underbrace{\sum_k \Delta_k a_k^{\dagger} a_k}_{H_{\text{B}}} \underbrace{-i \sum_{ik} g_k (\sigma_i^+a_k^\dagger - \sigma_i^-a_k)}_{H_{\text{I}}^{\text{odd}}}
    \end{aligned}
    \label{eqn:even/odd}
\end{equation}

Using expression~\ref{eqn:even/odd}, the lowest order term in the Magnus expansion is easily calculated

\begin{equation}
    \begin{aligned}
        \bar H^{(0)} &= \frac{1}{2\tau}\int_0^{2\tau}dtH(t) =\frac{1}{2\tau}\left(\int_0^{\tau}dtH_{\text{even}}+\int_\tau^{2\tau}dtH_{\text{odd}}\right)\\
        &=\sum_{k}\Delta_{k} a_{k }^\dagger a_{k } \nonumber -i\sum_{ik} \frac{g_{k}}{2}(\sigma_i^+ + \sigma_i^-) \left( a^\dagger_{k} - a_{k}\right)
    \end{aligned}
    \label{eqn:zeroth_order}
\end{equation}

\noindent which is equivalent to the many-emitter Weisskopf-Wigner Hamiltonian in the resonant case up to a factor of 2 in the interaction term. Thus, to zeroth order, the pulse sequence approximately cancels atomic dephasing after a full $2\tau$ cycle. 

It is rather straightforward to calculate the higher order terms by utilizing the standard commutation relations $[\sigma_i^+,\sigma_j^-] = \delta_{ij}\sigma_i^z$, $[\sigma_i^z,\sigma_j^\pm] = \pm2\delta_{ij}\sigma_i^\pm$, $[a_k,a_{k'}^\dagger]=\delta_{kk'}$,  $[a_k^\dagger a_k,a_{k'}^\dagger]=\delta_{kk'}a_k^\dagger$ and $[a_k^\dagger a_k,a_{k'}]=-\delta_{kk'}a_k$ and the identity $[AB,CD] = A[B,C]D + AC[B,D] +[A,C]DB + C[A,D]B$. The relevant Hamiltonian-term commutators are given by

\begin{equation}
    \begin{aligned}
        [H_A,H_I^{\text{odd}}]  &= -i\sum_{ik}\Delta_ig_k(\sigma_i^+a_k^\dagger+\sigma_i^-a_k)\\
        [H_B,H_I^{\text{odd}}]  &= -i\sum_{ik}\Delta_kg_k(\sigma_i^+a_k^\dagger+\sigma_i^-a_k)\\
        [H_I^{\text{even}}, -H_A]  &= i\sum_{ik}\Delta_ig_k(\sigma_i^-a_k^\dagger+\sigma_i^+a_k)\\
        [H_I^{\text{even}}, H_B]  &= i\sum_{ik}\Delta_kg_k(\sigma_i^-a_k^\dagger+\sigma_i^+a_k)\text{, and}\\
        [H_I^{\text{even}}, H_I^{\text{odd}}]  &= \sum_{ik}g_k\left(\sigma_i^z\sum_{k'}g_{k'}(a_k^\dagger a_{k'}^\dagger-a_k a_{k'})+\sum_jg_k(\sigma_i^+\sigma_j^+-\sigma_i^-\sigma_j^-)\right).
    \end{aligned}
    \label{eqn:hamiltonian-commutators}
\end{equation}

The commutators in Eqns.~\ref{eqn:hamiltonian-commutators} are then used to find the full even/odd Hamiltonian commutator

\begin{equation}
    \begin{aligned}
        [H_{\text{even}},H_{\text{odd}}] &= [H_A,H_I^{\text{odd}}]+[H_B,H_I^{\text{odd}}]+[H_I^{\text{even}},-H_A]+[H_I^{\text{even}},H_B] +[H_I^{\text{even}},H_I^{\text{odd}}]\\
        &=-i\sum_{ik}g_k^2\bigg\{\frac{\Delta_i+\Delta_k}{g_k}(\sigma_i^+-\sigma_i^-)\left(a_k^\dagger-a_k\right)+i\sum_{k'}\frac{g_{k'}}{g_k}\sigma_i^z\left(a_k^\dagger a_{k'}^\dagger-a_k a_{k'}\right) + i\sum_j(\sigma_i^+\sigma_j^+-\sigma_i^-\sigma_j^-)\bigg\}
    \end{aligned}
    \label{eqn:commutators}
\end{equation}

The next lowest (first) order term in the Magnus expansion is then

\begin{equation}
    \begin{aligned}
        \bar H^{(1)} &= \frac{1}{4i\tau}\int_0^{2\tau}dt\int_0^{t}dt'[H(t),H(t')]\\
        &=\frac{1}{4i\tau}\bigg(\int_0^{\tau}dt\int_0^{t}dt'[H_{\text{even}},H_{\text{even}}] +\int_\tau^{2\tau}dt\int_0^{\tau}dt'[H_{\text{odd}},H_{\text{even}}] +\int_\tau^{2\tau}dt\int_\tau^{2\tau}dt'[H_{\text{odd}},H_{\text{odd}}]\bigg)\\
        &=\frac{\tau}{4}\sum_{ik}g_k^2\bigg\{\frac{\Delta_i+\Delta_k}{g_k}(\sigma_i^+-\sigma_i^-)\left(a_k^\dagger-a_k\right)+i\sum_{k'}\frac{g_{k'}}{g_k}\sigma_i^z\left(a_k^\dagger a_{k'}^\dagger-a_k a_{k'}\right) + i\sum_j(\sigma_i^+\sigma_j^+-\sigma_i^-\sigma_j^-)\bigg\}.
    \end{aligned}
    \label{eqn:first_order}
\end{equation}

Higher order terms are marked by increasing powers of $\tau$. Namely, the $n^{\text{th}}$ order term in the Magnus expansion carries a factor of $\tau^n$. We thus find that to zeroth order, the net dephasing of the atoms averages to zero via the pulse protocol after a full $2\tau$ pulse cycle and that higher order terms vanish with increasingly rapid pulse rates.

%\bibliographystyle{apsrev4-1}
%\bibliography{References}
\end{document}